\documentclass[12pt]{article}
\usepackage{epsfig}

\usepackage{graphicx}
\usepackage{cancel}
\usepackage{amssymb,amsmath}
\def\harr#1#2{\smash{\mathop{\hbox to .3in{\rightarrowfill}}
 \limits^{\scriptstyle#1}_{\scriptstyle#2}}}

\def\s2{\frac{1}{\sqrt2}}

\def\beqa{\begin{eqnarray}}
\def\eeqa{\end{eqnarray}}

\def\Dsl{\,\raise.15ex\hbox{/}\mkern-13.5mu D} %can be subscripted
\def\d3{d^3}

\newcommand{\be}{\begin{equation}}
\newcommand{\ee}{\end{equation}}
\newcommand{\beq}{\begin{eqnarray}}
\newcommand{\eeq}{\end{eqnarray}}

\usepackage{amssymb,amsmath}

\usepackage{mathrsfs}%%%%%%%%%%%%%%% para la letra P de paridad

\def\be{\begin{equation}}
\def\ee{\end{equation}}
\def\beqa{\begin{eqnarray}}
\def\eeqa{\end{eqnarray}}

\begin{document}

\begin{center}
\Large{\bf Emergent Kalb-Ramond fields from a dimer model}
\vspace{0.5cm}

\large Luis Lozano\footnote{e-mail address:
{\tt llozano@fis.cinvestav.mx}}, Hugo Garc\'ia-Compe\'an\footnote{e-mail address: {\tt
compean@fis.cinvestav.mx}}

{\small \em Departamento de F\'{\i}sica, Centro de
Investigaci\'on y de Estudios Avanzados del IPN}\\
{\small\em P.O. Box 14-740, CP. 07000, M\'exico D.F., M\'exico}
\vspace*{0.5cm}

\vspace*{1.5cm}
\end{center}

\begin{abstract}
The emergence of a Kalb-Ramond field and string charge in the
lattice is discussed. The local bosonic model with rotor variables placed
on the faces of a cubic lattice is considered. The coupling model
consisting of the Maxwell fields and the Kalb-Ramond field is given.
This construction naturally incorporates the emerging coupling between both
gauge and string fields. In the process, an object that resembles to a D-brane
on the lattice is introduced.

\vskip 1truecm

\end{abstract}

\bigskip

\newpage

%%%%%%%%%%%%%%%%%%%%%%%%%%%%%%%%%%%%%%%%%%%%%%%%%%%%%%%%%%%%%%%%%%%%%%%%%%%%%%%%%%%%%%
%%%%%%%%%%%%%%%%%%%%%%%%%%%%%%%%%%%%%%%%%%%%%%%%%%%%%%%%%%%%%%%%%%%%%%%%%%%%%%%%%%%%%%
%%%%%%%%%%%%%%%%%%%%%%%%%%%%%%%%%%%%%%%%%%%%%%%%%%%%%%%%%%%%%%%%%%%%%%%%%%%%%%%%%%%%%%
\section{Introduction}
\label{sec:intro}

Recently a great deal of work has been done in the subject of the
topological phases of matter and topological order (see for instance
\cite{bookWen,Witten:2015aoa}). Moreover, it is now
believed that the diverse types of matter and its interactions would
be originated due the existence of a system formed by quantum bits
of information (see \cite{bookWen,Levin:2004js,Wen:2017usd} and
references therein). Thus, the matter can be regarded as an emergent
object coming from diverse ways of organization of some few degrees
of freedom in local lattice bosonic models. In this context, there is a
proposal in which electrons and photons and their interaction can
emerge from qubits in a string-net liquid. Thus electrons and photons
can be viewed as collective models of a string-net model
\cite{Levin:2004js}.

Beyond the model that gives rise to electrodynamics, in turn,
diverse string-net liquids can lead to different types of gauge
bosons and fermions with more general properties. Thus, it was
possible to find gauge bosons and fermions that behave as gluons and
quarks from an appropriate string-net local qubit model. That means
that QCD can be obtained as an emergent theory
\cite{Wen:2003ym,Levin:2004mi}. Later, interesting generalizations
to theories of gauge fields and massless fermions in any dimension
and for any gauge group were also constructed \cite{Levin:2004mi}.
In particular, the standard model of particles can be stated in this
context. However, the prediction of this description implies the existence
of discrete groups, which can be interpreted as cosmic strings in a very
early Universe.

On the other hand, superstring theory is a theory from which it is
possible as well to incorporate gauge bosons and fermions in a very
different way (see, for instance \cite{Green:2012oqa,
Polchinski:1998rq,1}. This procedure requires the introduction of
the idea of compactification. This roughly speaking implies to make
compact and small the extra dimensions in order to get the theory in
four dimensions. These considerations have been, by themselves,
problematic and the constructions of sensible field theories
requires to avoid the swampland, (for a review see for instance,
\cite{Palti:2019pca, Yamazaki:2019ahj}). Thus it seems natural to
search for other alternatives.

In the 90's with the advent of string dualities and the arising of
M-theory, there were some proposals involving the possible origin of
the fundamental strings and their properties, as derived objects
from more fundamental degrees of freedom. These degrees of freedom
were the non-perturbative objects known as D0-branes. A gas of $N$
of these D0-branes is, under certain considerations, described by
Matrix $N \times N$ quantum mechanics known as Matrix Theory
\cite{Banks:1996vh}. In AdS/CFT correspondence
\cite{Aharony:1999ti}, the gravitational fields are also emergent.
Gravity also can be considered an emergent interaction in matrix
models of gravity \cite{DiFrancesco:1993cyw}. For a review of
different aspects of emergent gravity, see \cite{Carlip:2012wa}.
Thus in superstring theories not only fermions and gauge field
emerges but in closed strings gravity arises at low energies.

The claim of deriving theories for a few degrees of freedom
localized in some region of spacetime is more general in the sense
that not only the diverse types of matter can be derived from the
local models of qubits. The idea is that the gravitational degrees
of freedom, and moreover spacetime by itself, might be obtained in
this way.

With the arrival of many new techniques from Condensed Matter
Physics, many efforts have been done to obtain gravitons and soft
gravitons as emergent particles from lattice models \cite{5,3,14},
starting from a symmetric rank-2 tensors immersed on the vertices
(diagonal terms) and on the faces of the lattices (off-diagonal
terms). Higher-rank symmetric tensors generalization of these works
were also proposed in Refs. \cite{7,8,9}.

Thus, in the present article we propose a local bosonic model
consisting in regarding the fundamental string and some of its
properties, as the Kalb-Ramond charge, as emergent objects from a
local lattice model. We will work with an anti-symmetric rank-2
tensor (see \cite{1,2} for a review of Kalb-Ramond fields) and, in
particular, we obtain emergent Kalb-Ramond fields from a lattice
model. With this purpose, we first introduce the model for
electromagnetism, and then we combine both models to obtain the
coupling of the Kalb-Ramond field potential to the string charge and
to the electric field. Earlier lattice models incorporating
Kalb-Ramond fields were proposed in \cite{Rey:1989ti,Rey:2010uz}. In
\cite{Rey:1989ti} a Higgs mechanism for the Kalb-Ramond fields is
proposed by coupling them to a string that eventually condensates.
Moreover in Ref. \cite{Rey:2010uz} a non-abelian tensor gauge theory
is implemented in the cubic lattice through the consideration of
Chan-Paton colors in each boundary link.

This article is organized as follows, in Section \ref{sec:prel} we
give some preliminary material concerning some facts about the
Kalb-Ramond field, the Maxwell field and their coupling. We also
revisited the photon model and the partition function. Section
\ref{sec:KBM} is the main part of our paper and it is devoted to
propose our model of emerging Kalb-Ramond field and the string
charge. Moreover, in this section we give also the lattice model of
the emerging coupling of the mentioned fields, which requires the
introduction of the idea of a D-brane in the lattice. Finally, in
Section \ref{sec:final} we give our final remarks.

%%%%%%%%%%%%%%%%%%%%%%%%%
%%%%%%%%%%%%%%%%%%%%%%%%%
\section{Preliminaries}
\label{sec:prel}

In the present section we give some preliminaries for Section \ref{sec:KBM}.
Here we will introduce the notation and conventions we will follow
in this article. We start by reviewing the field theory of the
Kalb-Ramond field including its sources \cite{9,2,Rey:1989ti}.

%%%%%%%%%%%%%%%%%%%%%%%%%
%%%%%%%%%%%%%%%%%%%%%%%%%
\subsection{String charge density}
\label{subsec:SCD}

We first review the string charge by introducing the antisymmetric
Kalb-Ramond field potential $ \mathcal{A}_{\mu \nu} =
-\mathcal{A}_{\nu \mu}$ on a $(3+1)$-dimensional Minkowski spacetime
and its associated field strength $\mathcal{F}_{\mu \nu \rho}$ given
by \be \mathcal{F}_{\mu \nu \rho} = \partial_\mu \mathcal{A}_{\nu
\rho} +
\partial_\nu \mathcal{A}_{\rho \mu}+ \partial_\rho \mathcal{A}_{\mu
\nu}. \label{one} \ee These fields have a great similarity with the
Maxwell gauge field potential $A_\mu $ and the electromagnetic field
strength $F_{\mu \nu}$ (note the ranks). In the electromagnetic
theory, the electric current $j^k$ (one index $k=1,2,3$) and the
electric charge density $q (= j^0)$ appear in \be \partial_\nu
F^{\mu \nu} = j^\mu. \label{two} \ee

For the Kalb-Ramond field strength we have \be {1 \over \kappa^2}
\partial_\rho \mathcal{F}^{\mu \nu \rho} = j^{\mu \nu},  \label{three} \ee where
$\kappa$ is a constant needed to keep the units, and $j^{\mu \nu}$
(two indices) is an antisymmetric tensor ($j^{\mu \nu} = -j^{\nu
\mu}$). The components $j^{0 k} = \vec{j}^0$ are called the
Kalb-Ramond charge density, or for simplicity, the string charge
density. It satisfies $\nabla \cdot \vec{j}^0 = 0,$ and in the case
of static strings we have $j^{i k} = 0,$ which is the case we adopt
in our lattice model, and only $j^{0 k}$ will be non-vanishing.

For static strings, we have to consider that $\partial_\rho
\mathcal{F}^{i k \rho} = 0,$ and also that \be \partial_\ell
\mathcal{F}^{0 k \ell} = \kappa^2 j^{0 k}. \label{four}\ee There is
a canonical conjugate variable $\Pi^{k \ell}$ to the string field
potential $\mathcal{A}^{k \ell}$ which can be obtained as $\Pi^{k
\ell} = \mathcal{F}^{0 k \ell}$ (see the \ref{subsec:HAM} and \cite{2}).
We can also introduce a vector $\vec{B}_{\mathcal{\mathcal{F}}}$
field related to the Kalb-Ramond field strength by \be
\mathcal{F}^{0 k \ell} = \varepsilon^{k \ell m} B_{\mathcal{F} m}.
\label{five} \ee It is called the field strength dual to
$\mathcal{F}$, and joining these last equations we obtain \be
\varepsilon^{k \ell m}
\partial_\ell B_{\mathcal{F} m } = \kappa^2 j^{0 k}, \label{six} \ee which is
like the Ampere's law but for the string charge densities.

%%%%%%%%%%%%%%%%%%%%%%%%%

\subsection{The Hamiltonian}
\label{subsec:HAM}

Now we turn to the Hamiltonian formulation of the free static
Kalb-Ramond field strength (see \cite{1,2,Rey:1989ti}), then first we
have to check the Lagrangian density \be \mathcal{L} = - {1 \over 6
\kappa^2} \mathcal{F}^{\mu \nu \rho} \mathcal{F}_{\mu \nu \rho},
\label{seven} \ee which is invariant under the gauge transformations
$ \mathcal{A}_{ij} \to \mathcal{A}_{ij} + \partial_i f_j -
\partial_j f_i  $ since we are working in the static case. The
canonical momentum conjugate to $ \mathcal{A}_{i j} $ is found to be
(check \cite{2} for details), \be \Pi^{i j} = \dot{\mathcal{A}}^{i
j} +
\partial^i \mathcal{A}^{j 0} + \partial^j \mathcal{A}^{0 i}. \label{eight}\ee
This is a two-rank antisymmetric tensor that satisfies $
\dot{\Pi}^{0 i} = \partial_k \Pi^{k i}, $ and the weak constraints
\be \Pi^{0 i} \approx 0,  \qquad \qquad \partial_i \Pi^{i k} \approx
0. \label{nine} \ee Thus to make the Hamiltonian density \be
\mathcal{H} = {1 \over 4} \Pi^{i j} \Pi_{i j} + {1 \over 2}
\mathcal{F}^{1 2 3} \mathcal{F}_{1 2 3} + \mathcal{A}_{j 0}
\partial_i \Pi^{i j} - \dot{\mathcal{A}}_{0 j} \Pi^{0 j}.  \label{ten}\ee

The first term is analogous to the term $E^2$ for electrodynamics,
and the second term for $B^2.$ They can be put together in the
Hamiltonian as one term like $\propto \mathcal{F}_{i j k}^2$ The
third and fourth term have been added to the Hamiltonian, so we can
modify them to fit the lattice model keeping them as constraints. As
can be observed from the last section, the term $\partial_i
\Pi^{ij}$ is the string charge $-\kappa^2 j^{0j},$ so if there is a
term with $\mathcal{A}_{j0}$ as a coefficient, it has to be
interpreted as the string charge. Depending on the gauge chosen,
this term can be taken as $0,$ but we leave that for now.

%%%%%%%%%%%%%%%%%%%%%%%%%

\subsection{Branes}
\label{subsec:BRA}

We would like to introduce a little about D-branes, which are going to
be helpful in the section in order to have a better understanding of how
the couplings occur. In the continuous case (see \cite{10}), we have that
the open strings are free on the space (in our case a 3-dimensional
space), but couple to D-branes on their boundaries or endpoints. These
objects are D-dimensional bodies that live also free on the space, as a
matter of fact, a 0-brane is a point, and a 1-brane is a string. In this way,
we can observe that a 2-brane is membrane, which is a 2-dimensional
object, and so on for following dimensions.

\begin{figure}
%\begin{minipage}[t]{8cm}
\begin{center}
\includegraphics[scale=0.6]{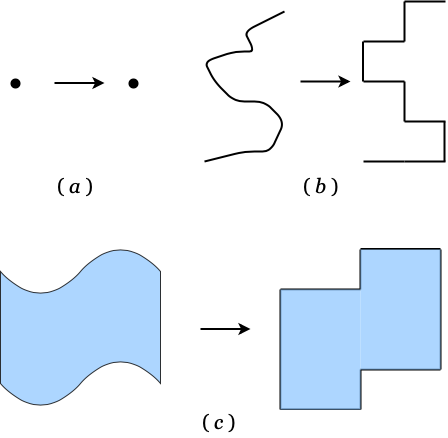}
\caption[Short caption for figure 1]{\label{Figure1} {\scriptsize Continuous and lattice D-branes. a) shows how a 0-brane is put on the lattice, b) shows a 1-brane on the lattice, and c) a 2-brane.}}
\end{center}
%\end{minipage}
\end{figure}

\begin{figure}
%\begin{minipage}[t]{8cm}
\begin{center}
\includegraphics[scale=0.6]{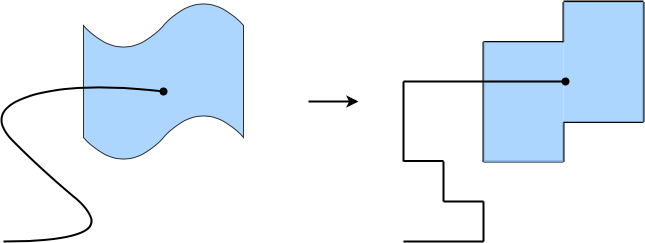}
\caption[Short caption for figure 1]{\label{Figure2} {\scriptsize Continuous and lattice strings attached to 2-branes }}
\end{center}
%\end{minipage}
\end{figure}

Objects like these, can be represented on a lattice by restricting
the space on which one is allowed to work inside the brane.
Different examples are shown on Figure \ref{Figure1}. Also, it is important to
have in mind how a string couples to a D-brane, as can be seen in
Figure \ref{Figure2}, where one of the ends of a string is attached to a
2-brane. Closed strings do not attach to D-branes, but open strings attach
to D-branes from both sides, it can be to the same D-brane or to a different
one. For the coupling model, we will consider this and restrict to the case in
which the string carries Kalb-Ramond charge, and only electric fields will be
treated on the D-branes.

%%%%%%%%%%%%%%%%%%%%%%%%%%%%%%%%%%%%%%%%%%%%%%%%%%%%%%%%%%%%%%%%%%%%%%%
%%%%%%%%%%%%%%%%%%%%%%%%%%%%%%%%%%%%%%%%%%%%%%%%%%%%%%%%%%%%%%%%%%%%%%%

%%%%%%%%%%%%%%%%%%%%%%%%%%%%%%%%%%%%%%%%%%%%%%%%%%%%%%%%%%%%%%%%%%%%%%%
%%%%%%%%%%%%%%%%%%%%%%%%%%%%%%%%%%%%%%%%%%%%%%%%%%%%%%%%%%%%%%%%%%%%%%%

%%%%%%%%%%%%%%%%%%%%%%%%%

\subsection{Couplings}
\label{subsec:CPL}

Now, we want to present the different couplings we will encounter in
the lattice model, beginning with the couples between a charge and
its corresponding field, with the final objective of coupling the
electric field with the string field (or Kalb-Ramond field).

So first, we remind the coupling between the gauge field potential
$A_\mu$ and the electric charge density vector $j^\mu$, which is
given by \be A_\mu j^\mu \label{eleven} \ee in the Hamiltonian
density. The conservation of the electric charge density
$\partial_\mu j^\mu = 0$ is implied by the gauge invariance under
the transformation \be A_\mu = A_\mu +
\partial_\mu f, \label{twelve} \ee as can be seen from the variation of the
coupling action (see \cite{1}). Here $f$ is an arbitrary scalar
field.

For the Kalb-Ramond charge and field we have the general form of the
coupling as \be \mathcal{A}_{\mu \nu} j^{\mu \nu}, \label{tteen} \ee
and, as in the electromagnetic case, invariance under gauge
transformation \be \mathcal{A}_{\mu \nu} \to \mathcal{A}_{\mu \nu} +
\partial_\mu \lambda_\nu - \partial_\nu \lambda_\mu, \label{ft}\ee
where the $\lambda_{\mu}$'s are arbitrary vector fields, implies the
conservation of string charge $\partial_\mu j^{\mu \nu} = 0$.

When the string is open, there is another type of coupling on the
boundaries given between the Kalb-Ramond field and the electric
field. For this matter, first we have to split the coordinates in:
coordinates normal or perpendicular to the electric field
($\mu_\perp$), and coordinates along the electric field or parallel
to it ($\mu_\parallel$), as $\mu = (\mu_\perp, \mu_\parallel)$.
Following this, we also have to couple the gauge transformations as
\be \mathcal{A}_{\mu \nu} \to \mathcal{A}_{\mu \nu} + \partial_\mu
\lambda_\nu - \partial_\nu \lambda_\mu, \label{fft} \ee \be
A_{\mu_\parallel} \to A_{\mu_\parallel} - \lambda_{\mu_\parallel},
\label{st} \qquad \quad \quad \ee where the term added for the
Maxwell gauge transformation is equivalent to the equation
(\ref{twelve}).

In order to keep the string action and the Maxwell action invariant,
we have to include the invariant quantity \be -{1 \over 4}(F+
\mathcal{A})^{\mu_\parallel \nu}(F + \mathcal{A})_{\mu_\parallel
\nu} \label{sst} \ee in the Hamiltonian, which, by expansion, gives
rise to the term \be -F^{0 k} \mathcal{A}_{0 k}. \label{et}\ee This
is the coupling between the electric field $F^{0 k} = E^k,$ which
takes the place of the string charge away from the string, and the
string field potential $\mathcal{A}^{0 k}$ (see \cite{1,Rey:1989ti}).

\begin{figure}
%\begin{minipage}[t]{8cm}
\begin{center}
\includegraphics[scale=0.6]{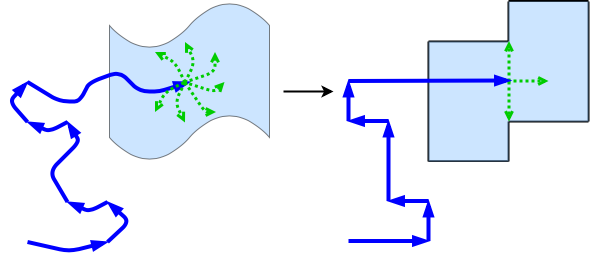}
\caption[Short caption for figure 1]{\label{Figure3} {\scriptsize Continuous and lattice string charge attached to 2-branes. The string charge is shown in blue, while the electric field is shown in green.}}
\end{center}
%\end{minipage}
\end{figure}

It has to be clear that the string charge couples to the string
field potential along the string (which is the place where the
string charge exists). On the other hand, the electric field couples
to the string field potential only through the D-brane to which the
string is attached, as can be seen in Figure \ref{Figure3}. It can also be
appreciated how the string charge goes along the string, and how the
electric field goes on the 2-brane.

%%%%%%%%%%%%%%%%%%%%%%%%%%%%%%%%%%%%%%%%%%%%%%%%%%%%%%%%%%%%%%%%%%%%%%%
%%%%%%%%%%%%%%%%%%%%%%%%%%%%%%%%%%%%%%%%%%%%%%%%%%%%%%%%%%%%%%%%%%%%%%%

%%%%%%%%%%%%%%%%%%%%%%%%%

\subsection{The photon model}
\label{sebsec: PHT}

Now we will present a dimer model that gives rise to photons as
emergent particles (there have been many other models, see for
example \cite{5,3,4,6,7}, and many generalizations to higher order
symmetric tensors, see \cite{5,3,8,9,10}).

Let's first, introduce the quantum dimer model (QDM see \cite{11})
with the simplest kinetic and potential energy terms written as \be
H_{QD} = \sum_{\square} \bigg\{ -T_1 (\mid = \rangle  \langle
\parallel \mid  + h. c.) + T_2 (\mid \parallel \rangle  \langle
\parallel \mid  +  \mid = \rangle  \langle = \mid) \bigg\},  \label{nt} \ee where the
summation runs over all the plaquettes ($\square$) of the lattice,
in which the plaquettes are the same as the faces.

\begin{figure}
%\begin{minipage}[t]{8cm}
\begin{center}
\includegraphics[scale=1.2]{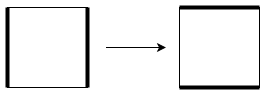}
\caption[Short caption for figure 1]{\label{Figure4} {\scriptsize The kinetic term flips the dimers as in equation (\ref{nt}).}}
\end{center}
%\end{minipage}
\end{figure}

In this model, the kinetic term $T_1$ flips pairs of
nearest-neighbor parallel dimers, which are links on the lattice
(see Figure \ref{Figure4}), and the potential term $T_2$ creates a
repulsion between them. This model has been used widely with different
purposes (see \cite{3,4,8,12,15,13}, but we will mostly follow the
meaning for the electrical part used in \cite{3} with a slight
variation in the notation as in \cite{16}.

%%%%%%%%%%%%%%%%%%%%%%%%%%%%%%%%%%%%%%%%%%%%%%%%%%%%%%%%%%%%%%%%%%%%%%%
%%%%%%%%%%%%%%%%%%%%%%%%%%%%%%%%%%%%%%%%%%%%%%%%%%%%%%%%%%%%%%%%%%%%%%%

We are going to define on each link $(i, \alpha)$ ($i=(i_x, i_y,
i_z)$ denotes the site in which the dimer begins, and $\alpha = x,
y, z$ the direction to which it grows) an number operator analog to
the electric field ${\widehat{E}}_{i \alpha}$ and its conjugate
angular phase operator analog to the potential field
${\widehat{A}}_{i \alpha}$ (as defined in \cite{3,4,15,16}), so
these variables satisfy $[{\widehat{A}}_{i \alpha}, {\widehat{E}}_{j
\beta}] = i \delta_{i j} \delta_{\alpha \beta}.$ It is important to
notice that the notation for the operators defined on the links
implies that $\widehat{E}_{i \alpha} = \widehat{E}_{i +
\widehat{\alpha}, - \alpha}.$

There is a constraint on the system that we have to impose on the
Hilbert space because it represents the discrete form of Gauss's law
for electric fields (see \cite{5,14,3}) \be \nabla_\alpha
{\widehat{E}}_{i \alpha} = 0, \label{tt} \ee where the symbol
$\nabla_\alpha$ means lattice differentiation or difference, and is
defined as $ \nabla_\alpha {\widehat{E}}_{i \beta} =
{\widehat{E}}_{i + \widehat{\alpha}, \beta} - {\widehat{E}}_{i
\beta}. $ Because of this constraint, the low-energy Hamiltonian has
to be invariant under the gauge transformation \be {\widehat{A}}_{i
\alpha} \to {\widehat{A}}_{i \alpha} + \nabla_{\alpha} f_i,
\label{to}\ee where $f_i$ is an arbitrary scalar field defined on
the sites. This last equation is the discrete version of the
equation (\ref{twelve}).

\begin{figure}
%\begin{minipage}[t]{8cm}
\begin{center}
\includegraphics[scale=0.6]{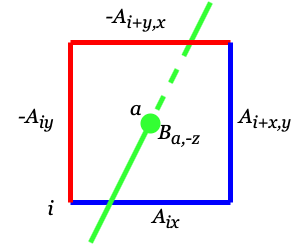}
\caption[Short caption for figure 1]{\label{Figure5} {\scriptsize The argument of the cosine function in equation (\ref{ttw}) is visible, and how it gives rise to a dual magnetic field $-B_{az}.$ The positive field potentials are shown in blue, while the negative ones are shown in red. The magnetic field generated is on the opposite direction and negative, shown in green.}}
\end{center}
%\end{minipage}
\end{figure}

With all this information, we have that the Hamiltonian for our
system is given by \be H_{e} = {K_1 \over 2} \sum_{i \alpha}
{\widehat{E}}_{i \alpha}^2 - K_2 \sum_{i \gamma} \cos(
\varepsilon_{\gamma \alpha \beta} \nabla_{\alpha} {\widehat{A}}_{i
\beta}). \label{ttw}\ee The notation employed is that of \cite{16}.
This Hamiltonian is the free three dimensional compact QED model
(see \cite{5}) with a deconfined photon phase (see \cite{3,17}). The
term with coefficient $K_1$ keeps a uniform density of dimers, and
the term with coefficient $K_2$ flips dimers around a plaquette. The
term $\varepsilon_{\gamma \alpha \beta} \nabla_{\alpha}
{\widehat{A}}_{i \beta}$ is shown in Figure \ref{Figure5} (see \cite{16}).

%%%%%%%%%%%%%%%%%%%%%%%%%%%%%%%%%%%%%%%%%%%%%%%%%%%%%%%%%%%%%%%%%%%%%%%
%%%%%%%%%%%%%%%%%%%%%%%%%%%%%%%%%%%%%%%%%%%%%%%%%%%%%%%%%%%%%%%%%%%%%%%

\begin{figure}
%\begin{minipage}[t]{8cm}
\begin{center}
\includegraphics[scale=0.5]{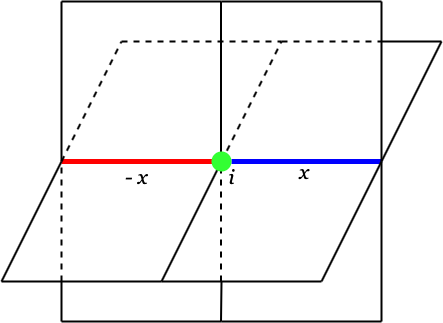}
\caption[Short caption for figure 1]{\label{Figure6} {\scriptsize A
defect at site $i$ when the Gauss' constraint is violated in the
$\widehat{x}$ direction. Again, blue is positive, red negative, and
green the charge. One can see that the 2 in the current on the
equation \ref{te} comes from the difference of the links. }}
\end{center}
%\end{minipage}
\end{figure}

In order to include currents and charges in our model, we have to
include a new phase angle operator $\widehat{\phi}_i$ on the sites,
and its conjugate number operator $\widehat{n}_i$. Also, there is an
important point about the Gauss' constraint, and it is that if it is
violated, some defects will be generated on the sites $i$ where
$\nabla_{\alpha} \widehat{E}_{i \alpha} \neq 0$, and these defects
carry charges of the gauge field potential $\widehat{A}_{i \alpha}$
(see Figure \ref{Figure6} and Ref. \cite{3}). The way in which the defects
couple to the gauge field potential is as a term in the Hamiltonian
as follows: \be H_{A.q} = -c_1 \sum_{\alpha} \cos(\partial_{\alpha}
\phi^{(q)} - A_{\alpha}). \label{tth}\ee

%%%%%%%%%%%%%%%%%%%%%%%%%%%%%%%%%%%%%%%%%%%%%%%%%%%%%%%%%%%%%%%%%%%%%%%
%%%%%%%%%%%%%%%%%%%%%%%%%%%%%%%%%%%%%%%%%%%%%%%%%%%%%%%%%%%%%%%%%%%%%%%

Considering the constraint (\ref{tt}) and the gauge transformation
(\ref{to}) we obtain a new Hamiltonian as: \be H_{\widetilde{e}} =
{K_1 \over 2} \sum_{i \alpha} {\widehat{E}}_{i \alpha}^2 - K_2
\sum_{i \gamma} \cos( \varepsilon_{\gamma \alpha \beta}
\nabla_{\alpha} {\widehat{A}}_{i \beta}) - K_3 \sum_i \widehat{n}_i
\widehat{A}_{i \tau} - K_4 \sum_{i \alpha} \cos(\nabla_{\alpha}
\widehat{\phi}_i - \widehat{A}_{i \alpha}). \label{tf} \ee The term
with coefficient $K_3$ is the coupling between the electric charge
and the electric field potential. It has to be clear that this term
is imposed by the Gauss' law in equation (\ref{tt}), in which the
term $\widehat{A}_{i \tau}$ can be seen as a Lagrange multiplier
defined on the vertices. The term with coefficient $K_4$ is the
gauge transformation of equation (\ref{tth}) defined on each link.
It is important to notice that for the gauge transformation inside
the cosine of the $K_4$ term to be valid in equation (\ref{tf}),
there has to be no current or $K_3 \to \infty$ as mentioned above,
and we work with the low-energy physics.

We now follow the line of Ref. \cite{16}, so we write the path
integral representation of the partition function by inserting the
eigenstates of $\widehat{E}_{i \alpha}$ at small imaginary time
intervals $\Delta \tau.$ The cosine term of $K_2$ can be replaced by
the Villain form approximation as \be \exp\bigg\{K_2 \Delta \tau
\cos (\varepsilon_{\gamma \alpha \beta} \nabla_{\alpha}
\widehat{A}_{i \beta})\bigg\} \to \sum_{\{B_{a \gamma} \}}
\exp\bigg\{- {B_{a \gamma}^2 \over 2 K_2 \Delta \tau} + i B_{a
\gamma} \varepsilon_{\gamma \alpha \beta} \nabla_{\alpha}
\widehat{A}_{i \beta}\bigg\}. \label{tf} \ee The corresponding to
$K_4$ is given by \be \exp\bigg\{K_4 \Delta \tau \cos
(\nabla_{\alpha} \widehat{\phi}_i - \widehat{A}_{i \alpha})\bigg\}
\to \sum_{\{ j_{i \mu} \}} \exp\bigg\{-{j_{i \mu}^2 \over 2 K_4
\Delta \tau} + i j_{i \alpha} (\nabla_{\alpha} \widehat{\phi}_i -
\widehat{A}_{i \alpha})\bigg\}, \label{ts} \ee where we have to keep
in mind that the Gauss' law along with the gauge transformation keep
this term as $0$.

Here, $B_{a \gamma}$ is an integer dual magnetic field defined on the
links $(a, \gamma)$ of the dual lattice (see Figure \ref{Figure5} and Ref.
\cite{18}). The relationship between the sites of the direct lattice (with
indexes $i, j, \dots,$) and the sites of the dual lattice (with indexes $a,
b, \dots,$) will not be taken into account explicitly, but it is well
understood from the example of the dual magnetic field that a
link in the dual lattice represents a face in the direct lattice,
and sites on the dual lattice are located at the center of the cubes
on the direct lattice and vice versa.

Furthermore, we define an integer dual electromagnetic tensor
$\widetilde{F}_{a \mu \nu}$ on the dual lattice (as a 3-dimensional
analog to that one in \cite{16}), with its row components as
$\widetilde{F}_{a x \nu} = (B_{a x}, 0, E_{i z}, -E_{i y}),$
$\widetilde{F}_{a y \nu} = (B_{a y}, -E_{i z}, 0, E_{i x}),$ and
$\widetilde{F}_{a z \nu} = (B_{a z}, E_{i y}, -E_{i x}, 0),$ where
$B_{a, -z}$ is the field generated as in Figure \ref{Figure5}. We also define
the electric current on the direct lattice as $j_{i \mu} = (-n_i,
k_{ix}, k_{iy}, k_{iz}),$ where the $k_{i \alpha}$ can be written in
terms of the variation of the defects $\widehat{\phi}$ between the
vertex $i$ and the vertex $i + \alpha$. Note that for both the dual
electromagnetic tensor, and the current, we use $\lambda, \mu, \nu, \rho =
\tau, x, y, z;$ and $\alpha, \beta, \gamma, \delta = x, y, z.$

We are now ready to obtain the following partition function after
working with the Villain form for the cosine terms as: \be
Z_{\widetilde{e}} = \sum_{\{\widetilde{F}_{a \mu \nu}, j_{i \mu} \}}
\exp\bigg\{-{e_1^2 \over 2} \widetilde{F}_{a \mu \nu}^2 + {g_1 \over
2} j_{i \mu} A_{i \mu}\bigg\}, \label{tseven} \ee restricted to \be
\varepsilon_{\lambda \mu \nu \rho} \nabla_{\mu} \tilde{F}_{a \nu \rho}
 + 2 j_{i \lambda} = 0, \label{te} \ee and \be \nabla_{\lambda }
 \widetilde{F}_{a \lambda \mu} = 0, \label{tn} \ee where the time interval
is chosen to give $e_1^2 = K_1 \Delta \tau = {1 \over K_2 \Delta \tau}$,
and $g_1 = K_3 \Delta \tau = {1 \over K_4 \Delta \tau}.$ It is also important
to observe that equations (\ref{te}) and (\ref{tn}) give rise to $\nabla_{\lambda}
j_{i \lambda} = 0,$ where in every case we are using all the subindices without
taking assumptions on the time differentiation, which can be taken as $0$
for simplicity.

Notice that the phases obtained in this case can be compared to
those obtained in Refs. \cite{3,4,16}, but the main difference is
that we are not working with holes because our Hamiltonian does not
include a term taking it into account, which means $0$ hole average,
as reflected on the partition function and the Gauss' constraint.

In order to give some solutions, we use the restrictions to obtain that
\be j_{i \lambda} = \varepsilon_{\lambda \mu \nu \rho} \nabla_{\mu}
a_{i \nu \rho}, \label{g1} \ee and \be \tilde{F}_{a \nu \rho} = \nabla_{\nu}
N_{a \rho} - 2a_{a \nu \rho}. \label{g2}\ee These equations are like those of Ref.
\cite{16}, but on 3 dimensions (+ 1 if the time is taken into account as
evolution). It can be observed that the current $j_{i \lambda}$ is generated by
a field on the faces of the direct lattice $a_{i \nu \rho}$, which is an integer field
on the links of the dual lattice.

The important phase to consider for the photon model is when the term
$g_1 \to \infty$we can see that either the the gauge $a$ or the field potential
$A$ have to be $0$, which means that there is no coupling. On the other
side, the other term is always giving rise to a stable phase with a maximum
on the energy when equation (\ref{g2}) is $0,$ which implies that the creation
of current is directly given by the integer $N$ variable. In Ref. \cite{16} it is
called the Higgs scalar and is defined on the vertices, but on our case it is
defined on the links, so it can be taken as a scalar number quantity that gives
rise to the electric and magnetic fields as this Higgs quantity fluctuates. The
term $g_1$ serves as an order parameter between these two phases as
$g_1 > 0$ on the phase with no couplings, to the phases on which there are
couplings and currents with $g_1 < 0.$

%%%%%%%%%%%%%%%%%%%%%%%%%

\section{ The Kalb-Ramond model}
\label{sec:KBM}

Turning now to the model for the string charges which is the same
lattice of the model above for electromagnetism, but now we define
different variables. First, we introduce the integer number
operators or boson numbers as $\widehat{n}_{i  \alpha \beta}$ on
each face $i + {\widehat{\alpha} \over 2} + {\widehat{\beta} \over
2}$ (with the constraint $\alpha \neq \beta$). Conjugate to these
variables, we have the phase angle operators $\widehat{\theta}_{i
\alpha \beta}$, defined on the same faces, related to the boson
($\widehat{n}_{i \alpha \beta}$) creation operators by
$\widehat{b}_{i \alpha \beta} \propto e^{-i \widehat{\theta}_{i
\alpha \beta}}$ and by the commutation relations on the face $i$ by
\be [\widehat{n}_{i \alpha \beta}, \widehat{\theta}_{i \gamma
\delta}] = i \delta_{\alpha \gamma} \delta_{\beta \delta}.
\label{ttt} \ee

We also define an antisymmetric tensor $\widehat{\Pi}_{i \alpha
\beta} = \varepsilon_{\alpha \beta} (\widehat{n}_{i \alpha \beta} -
\bar{n})$ and its conjugate antisymmetric tensor
$\widehat{\mathcal{A}}_{i \alpha \beta} = \varepsilon_{\alpha \beta}
\widehat{\theta}_{i \alpha \beta},$ where both are defined purely on
the faces because of the antisymmetry ($\widehat{\Pi}_{\alpha \beta}
= - \widehat{\Pi}_{\beta \alpha}$). The average density of bosons
per site and face is written as $\bar{n}$.

The Hamiltonian of our system is written as follows \be H_k = H_t +
H_u + H_0, \label{3o} \ee with \be H_t = -t \sum_{i \alpha \beta
\gamma \delta} \widehat{b}_{i \alpha \beta}^{\dagger} \widehat{b}_{i
\gamma \delta}, \label{3t} \ee \be H_u = u \sum_{\alpha \beta}
(\widehat{n}_{i \alpha \beta} - \bar{n})^2, \label{3tt} \ee \be H_0
\mid_{(i, x)} = U\bigg(\widehat{n}_{i + {\widehat{x} \over 2} +
{\widehat{z} \over 2},zx} + \widehat{n}_{i + {\widehat{x} \over 2} -
{\widehat{z} \over 2},zx} - \widehat{n}_{i + {\widehat{x} \over 2} +
{\widehat{y} \over 2},xy} - \widehat{n}_{i + {\widehat{x} \over 2} -
{\widehat{y} \over 2},xy}\bigg)^2. \label{3f}\ee

\begin{figure}
%\begin{minipage}[t]{8cm}
\begin{center}
\includegraphics[scale=0.5]{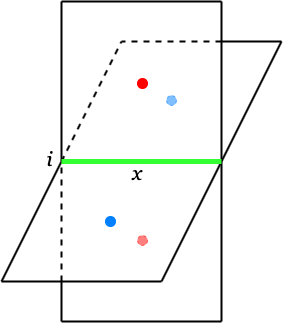}
\caption[Short caption for figure 1]{\label{Figure7} {\scriptsize The dimer $(i,x)$ (in green) is created by the interaction of the positive boson numbers (in blue) and the negative ones (in red) as the constraint $H_0$ dictates.}}
\end{center}
%\end{minipage}
\end{figure}

The first term $H_t$ is a hopping term between the bosons that are
on the nearest neighbor faces, the second term $H_u$ is a repulsive
interaction between them, and the last term $H_0 \mid_{(i,x)}$ is
the component along the link in the $x$ direction beginning at the
site $i$ relating the bosons that are on the faces touching that
link. For the term $H_0$, the terms along the $y$ and the $z$
direction are defined similarly (see Figure \ref{Figure7} for a reference).

%%%%%%%%%%%%%%%%%%%%%%%%%%%%%%%%%%%%%%%%%%%%%%%%%%%%%%%%%%%%%%%%%%%%%%%
%%%%%%%%%%%%%%%%%%%%%%%%%%%%%%%%%%%%%%%%%%%%%%%%%%%%%%%%%%%%%%%%%%%%%%%

This Hamiltonian is the lattice version of the equation (\ref{ten})
with the modifications mentioned below it. The $H_0$ term is a local
constraint like the one of electromagnetism (similar to the one
exposed in \cite{5,3,4} for soft gravitons and linear gravity) which
is given by \be \nabla_{\alpha} \Pi_{\alpha \beta} = 0 \label{3five}
\ee (with summation over the repeated index). This can take us back
to the canonical momentum ($\Pi_{ij}$) conjugate to the string field
potential ($\mathcal{A}_{ij}$) in the continuous case, which had a
gauge transformation (\ref{ft})). In our lattice system, if we
impose $U \to \infty$, the constraint has to be kept equal to $0$,
and in the low-energy Hamiltonian, this constraint imposes the
Hamiltonian to be invariant under a gauge transformation \be
\widehat{\mathcal{A}}_{i \alpha \beta} \to \widehat{\mathcal{A}}_{i
\alpha \beta} + \nabla_{\alpha} \lambda_{i \beta} - \nabla_{\beta}
\lambda_{i \alpha}, \label{3s} \ee with the $\lambda_{i \alpha}$'s
arbitrary vector fields defined on the links.

In the low-energy physics, the $H_t$ term of the Hamiltonian is
compactified (see \cite{5,3,4}), and writing the $H_u$ with the
antisymmetric tensor $\Pi_{i \alpha \beta}$, we have the following
effective Hamiltonian \be H_{k_{eff}} = -\widetilde{t} \sum_{i
\alpha, \beta \neq \alpha, \gamma \neq \beta} \cos(K
\nabla_{[\alpha} \widehat{\mathcal{A}}_{i \beta \gamma]}) + u
\sum_{i \alpha, \beta \neq \alpha} \widehat{\Pi}_{i \alpha \beta}^2,
\label{3seven} \ee where the term inside the cosine is related to
the term $\sim \mathcal{F}^{123} \mathcal{F}_{123}$ by a constant
$K$ (a modification is shown below), and the summation is intended
along the face. $\widetilde{t}$ is a constant related to a high
order perturbation of $t/U$ derived in the compactification process.
Comparing our model to that of electromagnetism of the last section,
we have plaquettes instead of dimers for the terms inside the
cosine, visibly as in Figure \ref{Figure5} (see \cite{8}).

%%%%%%%%%%%%%%%%%%%%%%%%%%%%%%%%%%%%%%%%%%%%%%%%%%%%%%%%%%%%%%%%%%%%%%%
\begin{figure}
%\begin{minipage}[t]{8cm}
\begin{center}
\includegraphics[scale=0.5]{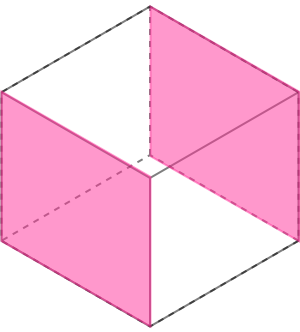}
\caption[Short caption for figure 1]{\label{Figure8} {\scriptsize The terms inside of the cosine of equation (\ref{3five}) are plaquettes instead of dimers.}}
\end{center}
%\end{minipage}
\end{figure}
%%%%%%%%%%%%%%%%%%%%%%%%%%%%%%%%%%%%%%%%%%%%%%%%%%%%%%%%%%%%%%%%%%%%%%%

We can observe that if the equation (\ref{3five}) is violated, we
obtain the string charge (Kalb-Ramond charge) as can be seen in the
continuous case in equations (\ref{four}) and (\ref{eight}). This
string charge is a defect and is taken by the string field potential
$\widehat{\mathcal{A}}_{i \alpha \beta}$. As can be observed from
the constraints (\ref{3tt}) and (\ref{3five}), and from the behavior
of the string charge (\ref{six}), we can see that these defects have
to travel along the links. To couple these defects to the string
field in a gauge invariant way, we need to consider the gauge
transformation (\ref{3s}) and observe that the defects can be added
by arbitrary vector fields working there, so we have the following
gauge invariant coupling term for the Hamiltonian (see \cite{3}) \be
H_{\mathcal{A} \cdot j} = -c_2 \sum_{\alpha, \beta}
\cos(\partial_{\alpha} \phi_{\beta}^{(j)} -
\partial_{\beta} \phi_{\alpha}^{(j)} - \mathcal{A}_{\alpha \beta}),
\label{3e} \ee where the $\phi_a^{(j)}$'s are interpreted as the
creation operators of the string charge, and the sum is along all
the links surrounding the face $\alpha \beta$.

In order to obtain a dimer model like the electromagnetic one, we
first have to define the variables required for equation (\ref{3e}).
In this way, we have the boson number operator $\widehat{n}_{i
\alpha}$ on the link $(i \alpha)$, as well as the conjugate operator
the phase angle operator $\widehat{\phi}_{i \alpha}$ defined
likewise. Then, we can work with the following description:
$$
H_{\widetilde{k}} = {K_5 \over 2} \sum_{i \alpha \beta}
\widehat{\Pi}_{i \alpha \beta}^2 - K_6 \sum_{i} \cos( \nabla_x
\widehat{\mathcal{A}}_{i y z})
$$
\be - K_7 \sum_{i \alpha}
\widehat{n}_{i \alpha} \widehat{\mathcal{A}}_{i \tau \alpha} - K_8
\sum_{i \alpha \beta} \cos(\nabla_{\alpha} \widehat{\phi}_{i \beta}
- \nabla_{\beta} \widehat{\phi}_{i \alpha} -
\widehat{\mathcal{A}}_{i \alpha \beta} ). \label{3n}\ee

The term with coefficient $K_5$ of equation (\ref{3n}) is kept as it
was, but the one with coefficient $K_6$ was modified as mentioned.
The term $K_7$ uses $\widehat{\mathcal{A}}_{i \alpha \tau}$ as a
Lagrange multiplier defined on the links, and we need $K_7 \to
\infty$ in order to keep the Gauss-like constraint. The last term
with coefficient $K_8$ is the modification for the lattice of
equation (\ref{3e}), implied from the $K_7$ restrictions.

We now will take the Villain form approximation in the path integral
representation for the $K_8$ term of equation (\ref{3n}). This will
result in the following process $$\exp\bigg\{K_8 \Delta \tau \cos
(\nabla_{\alpha} \widehat{\phi}_{i \beta} - \nabla_{\beta}
\widehat{\phi}_{i \alpha} - \widehat{\mathcal{A}}_{i \alpha
\beta})\bigg\}$$ \be \to \sum_{\{j_{i \mu \beta} \}} \exp\bigg\{-
{j_{i \mu \beta}^2 \over 2 K_8 \Delta \tau} + i j_{i \alpha \beta}
(\nabla_{\alpha} \widehat{\phi}_{i \beta} - \nabla_{\beta}
\widehat{\phi}_{i \alpha} - \widehat{\mathcal{A}}_{i \alpha
\beta})\bigg\}, \label{f} \ee where we have to take into account
that both terms will sum up to $0$ as well as we keep the Gauss-like
constraint (\ref{3five}) and the gauge transformation.

Now, we can proceed by defining the string field by components as
$\mathcal{F}_{i \tau \alpha \beta} = \Pi_{i \alpha \beta}$ and
$\mathcal{F}_{i \alpha \beta \gamma} = \nabla_{\alpha}
\mathcal{A}_{i \beta \gamma} + \nabla_{\beta} \mathcal{A}_{i \gamma
\alpha} + \nabla_{\gamma} \mathcal{A}_{i \alpha \beta},$ where we
are working with the eigenvalues of the operators. As in the
electromagnetic case, we can work with the Villain form
approximation for the last term of equation (\ref{3n}) to obtain the
partition function (see \cite{16}) \be Z_{\widetilde{k}} = \sum_{\{
\mathcal{F}_{a \mu \alpha \beta}, j_{i \mu \alpha} \}} \exp\bigg\{
-{e_2^2 \over 2} \mathcal{F}_{a \mu \alpha \beta}^2 + {g_2 \over 2}
j_{i \tau \alpha} \mathcal{A}_{i \tau \alpha}\bigg\}, \label{fo} \ee
restricted to \be \nabla_{\alpha} \mathcal{F}_{i \mu \alpha \beta} +
m j_{i \mu \beta} = 0, \label{ftwo}\ee with conserved string charge.
The index $\mu$ runs over $\tau, x, y, z$, and the other indices
only on the spatial dimensions. The term $m = 2, 4$ depending on
how many variables create the charge (see Figure \ref{Figure7}).
Only the $j_{i \tau \alpha}$ components survive for the string
charge, the other terms appear when we have a D-brane with
$D>1$ and with time evolution. The time interval is chosen to give
$e_2^2 = K_5 \Delta \tau = {1 \over K_6 \Delta \tau}$, and $g_2 =
K_7 \Delta \tau = {1 \over K_8 \Delta \tau}.$

We can make an analysis similar to the one on Ref. \cite{3}, but our
interest is the stable phase on the partition function of equation
(\ref{fo}). It is observed how the string charge $j_{i \tau \alpha}$
appears in the last step of the process as in the electromagnetic
case. The spatial components of this charge are canceled by the
constraint because we are working in a quasi-static case (see Ref.
\cite{1}). Also, it has to be clear that this phase is made only of
closed strings formed along the links of the direct lattice, at the
links where the string charge is defined.

Here we could also do the same analysis as we did for the photon
model, but it can be observed that the pure Kalb-Ramond field can
be seen as a photon model but worked on a dual lattice since all the
variables defined on the links for QED are now defined on the faces.
So pure electrodynamics and pure anti-symmetric fields are analogous
or "dual". A different analysis will be obtained when working on the
coupling model.

%%%%%%%%%%%%%%%%%%%%%%%%%

\subsection{ The coupling model} \label{subsec:CM}

Now, we can work on the coupling of the Kalb-Ramond charge with the
electric field, but there will be no electric charges or currents on
the system, so we will have to modify the term $K_4$ of the
Hamiltonian $H_{\widetilde{e}}.$ For this, we go back to equations
(\ref{fft}) and (\ref{st}), where we can see that the coupling is
through the potentials, and that the defects created will be along
the links. Following this, we have that the way in which these
defects couple to the electromagnetic field potential is: \be
H_{A.j} = -c_3 \sum_{\alpha_{\parallel}} \cos(\phi_{\alpha}^{(j)} +
A_{\alpha}), \label{fttt} \ee where the sum is along the coordinates
parallel to direction of the electric field. With its modification
to fit in the lattice, we obtain the $H_{e'}$ for our total
Hamiltonian which is \be H_{e'} = {K_{1'} \over 2} \sum_{i_b \alpha}
{\widehat{E}}_{i_b \alpha}^2 - K_{3'} \sum_{i_b \alpha}
\widehat{n}_{i_b \alpha} \widehat{E}_{i_b \alpha}, \label{ff} \ee where the
$K_{3'}$ is the term that couples the electric field with the string
charge. The coupling term with the electric field is only along the terms
parallel to the electric field, on the D-branes as \be H_{int} = -K_{4'} \sum_{i_b
\alpha_{\parallel}} \cos(\widehat{\phi}_{i_b \alpha_{\parallel}} +
\widehat{A}_{i_b \alpha_{\parallel}}. \label{ffive}), \ee where the vertices $i_b$
are the vertices that belong to the D-branes where the electric fields are
defined.

The Hamiltonian for the Kalb-Ramond charge and fields is kept
exactly as in equation (\ref{3n}). Then, the total Hamiltonian will
be \be H = H_{\widetilde{k}} + H_{e'} + H_{int}, \label{fs} \ee and, given
that we have the modification for the electric part, we will obtain
a Villain form approximation in the path integral representation for
the $K_{4'}$ term as \be \exp\bigg\{K_{4'} \Delta \tau
\cos(\widehat{\phi}_{i_b \alpha_{\parallel}} + \widehat{A}_{i_b
\alpha_{\parallel}})\bigg\} \to \sum{} \exp\bigg\{i \mathcal{A}_{i_b
\tau \alpha_{\parallel}} \widehat{\phi}_{i_b \alpha_{\parallel}} + i
\mathcal{A}_{i_b \tau \alpha_{\parallel}} \widehat{A}_{i_b
\alpha_{\parallel}}\bigg\}. \label{fseven} \ee

Here, the defect $\widehat{\phi}_{i_b \alpha_{\parallel}}$ can be
interpreted as the electric field, given the fact that away from the
open string, the string potential couples to the electric field
through a $\tau \alpha$ component. In this way, we obtain the
partition function as \be Z = \sum_{\{E_{i_b \alpha_{\parallel}},
\mathcal{F}_{a \mu \alpha \beta}, j_{i \mu \alpha} \}}
\exp\bigg\{-{e_1^2 \over 2} E_{i_b \alpha_{\parallel}}^2  - {e_2^2
\over 2} \mathcal{F}_{a \mu \alpha \beta}^2 + {g_1 \over 2} E_{i_b
\alpha_{\parallel}} \mathcal{A}_{i_b \tau \alpha_{\parallel}} + {g_2
\over 2} j_{i \tau \alpha} \mathcal{A}_{i \tau \alpha}\bigg\},
\label{fe} \ee restricted to \be \nabla_{\alpha} \mathcal{F}_{i
\mu \alpha \beta} + m j_{i \mu \beta} = 0, \label{fn} \ee and the
Maxwell equations without charges (see equations (\ref{te}) and
(\ref{tn})).

We have no magnetic field here because we are on a stationary case
with no electric currents and charges. The string field potential couples
to the electric field (on the D-brane where the electric field can be
defined) and to the string charge (only along the string). In this
stable phase, we have open strings attached to D-branes where
electric fields are defined up to short distances (because it costs
energy to keep electric fields), and we also have closed strings
only with string charge. The string charge may only have a value
different from $0$ for the coordinates $\tau \alpha.$

Now we see that the phases on the Hamiltonian for this coupling
model are diverse, but focusing on the different locations where these
couplings happen we can obtain the stable phases. First, we can see
a solution obtained from the restrictions for the electric fields as \be
E_{i_b \alpha_{\parallel}} = \varepsilon_{\alpha_{\parallel} \beta \mu \nu}
\nabla_{\beta} e_{i_b \mu \nu},
\label{g3} \ee where the electric field on the D-branes can be seen to be
generated from fluctuations of another integer variable $b.$ The current
has to be conserved, and given that it travels only along the strings we
have \be j_{i \tau \beta} = \varepsilon_{\beta \mu \nu \rho} \nabla_{\mu}
b_{i \nu \rho}, \label{g4}\ee where the term $b$ is an integer gauge field
that gives rise to the string charge. The difference between the string charge
and the electric field is where they are located, as is said on the subindices.
The string charge is only defined on the strings, and the electric field is only
defined on the D-branes. Finally, for the restriction on the equation (\ref{fn})
we obtain that \be \mathcal{F}_{i \mu \alpha \beta} = \nabla_{\mu}
N_{i \alpha \beta} - m \delta_{\tau \mu} b_{i \alpha \beta}, \label{g5}\ee
where the term $N$ can be seen as an integer defined on the faces of the
direct lattice.

We can observe that along the strings $e_1 \to \infty,$ $g_1 = 0,$ and
we will be working only with the string variables. Here, the important
phase is when $g_2 \to \infty,$ which is when there is no coupling, so
the Kalb-Ramond field potential and its conjugate field are given by the
fluctuations of the integer quantity $N$, which can be considered as the
Higgs variable on the faces. At the maximum for the Hamiltonian this term
has to be $0,$ so there can exist a balance between this fluctuations and the
gauge fields $b$ that generate the string charge, but this can only happen
when the term $g_2$ is finite or close to $0.$ When this happens, the current
is created by the fluctuations of the variable $b$ which also affects the creation
of the string fields.

The term $g_2$ can work as an order parameter that takes the system from a
stable phase with pure string fields $g_2 > 0,$ to a stable phase with couplings
between the string charges and the gauge potential fields $g_2 < 0$.

When we take into consideration the D-branes, we have that $g_2 = 0,$
where the electric fields are generated, as has been said, by the fluctuations
of an integer gauge field variable $b.$ Here, the
string fields are only generated by the fluctuations of the Higgs term $N$ when
$g_1$ is finite. As the term $g_1 \to \infty$ there is no coupling, so the $b$ does
not generate the electric field and only the string terms are generated.

%%%%%%%%%%%%%%%%%%%%%%%%%

\section{Final Remarks}
\label{sec:final}

In the present article we have proposed a model to obtain
electromagnetism $(3+1)$-dimensions from a dimer model. We have
found also its corresponding partition function, finding in the way
the dual electromagnetic tensor in 3+1 dimensions. The model is
assumed to be quasi-static in 3 spatial dimensions, but the
extensions to evolve on time have been marked along the construction
of the model. It is also important to remark that we used as the
source for the electric and magnetic fields the electric charges and
currents instead of the field potentials (see Refs. \cite{14,3,31}). This has to
be taken into account when calculating the correlation functions.

In addition, we have obtained Kalb-Ramond fields and charges from a
dimer model in 3+1 dimensions. We again assumed that the system is
quasi-static, but it has been marked through the construction how to
evolve on time and the places where the field potentials grow. Its
source has been taken to be the string charge, and because the model
is quasi-static, only the components that run along the strings
$j_{i \tau \alpha}$ can be different from $0$.

In the last section, we have been able to couple both models into
one 3+1-dimensional dimer model, this time taking as the source the
string charges. We have also pointed out that the coupling of the
string charges with the string field potentials has to be along the
strings (open or closed), while the coupling of the string field
potential with the electric field has to be on the D-branes attached
at the endpoints of the open strings.

In Ref. \cite{31}, the correlation functions were obtained using
Monte Carlo simulations. It would be interesting to apply these
methods within this context and compute some quantum observables. We
would like for the near future to search for a relation to the
results involving Kalb-Ramond fields \cite{Rey:1989ti,Rey:2010uz}.
Moreover, the model of the antisymmetric field studied in the
present article could be coupled to the symmetric model (as in Refs.
\cite{5,14,3}) as a solution to the problem of finding linearized
gravity as an emergent theory from lattice models. The resulting
model would be interesting in the study of gravitational models with
torsion \cite{Hammond:2002rm}. It is also of our interest to study
the possible relationship of our results in the context of
fermion-fermion duality \cite{Palumbo:2019tmg}. Some of these issues
will be reported elsewhere.

\newpage

%%%%%%%%%%%%%%%%%%%%%%%%%%%%%%%%%%%%%%%%%%%%%%%%%%%%%%%%%%%%%%%%%%%%%%%%%%
 \vspace{.5cm}
\centerline{\bf Acknowledgments} \vspace{.5cm}

Luis Lozano would like to thank CONACyT for a grant with CVU number 594425.

%%%%%%%%%%%%%%%%%%%%%%%%%%%%%%%%%%%%%%%%%%%%%%%%%%%%%%%%%%%%%%%%

\end{document}